# XFO: Toward Programming Rich Semantic Models

Robert B. Allen and Teryn K. Jones
rba@boballen.info, tkjones93@gmail.com

**ABSTRACT**

We have proposed that ontologies and programming languages should be more closely aligned. Specifically, we have argued that the Basic Formal Ontology (BFO2) has many features that are consistent with object-oriented analysis, design, and modeling. Here, we describe the eXtended Formal Ontology (XFO), a programming environment we developed to support semantic modeling. We then use XFO to implement a Traffic Light Microworld and discuss more complex applications.

## 1 INTRODUCTION

Upper ontologies define entity types that can be specialized and included in application ontologies. The Basic Formal Ontology (BFO2) [7] is a realist upper ontology that is grounded in Aristotelian principles and widely used in biomedicine. One way it is Aristotelian is in distinguishing Universals and Particulars. Particulars (instances) are portions of reality. Universals are abstractions of Particulars.

At the top level of BFO2, Entities are divided into Continuants and Occurrents. Continuants (3D) persist through time while Occurrents (4D) embody change. Continuants are further divided into Independent Continuants (e.g., Material Entities such as Objects) and Dependent Continuants (e.g., Qualities, Roles).

BFO2 also includes built-in relationships. For instance, the "Participates_In" relationship links Independent Continuants to Occurrent Processes.

BFO2 is a pure upper ontology. It focuses on identifying types of entities and allows domain applications to extend the entity types, but does not include domain applications. In [5], we proposed developing a semantic Model Layer to describe the interaction of BFO2 Entities. In this paper, we describe a programming environment to implement that Model Layer. There are many advantages in having computational tools for building a Model Layer. Ultimately, the tools could support the large-scale, high-structured direct representation of scientific research reports [4] and descriptions of history [6].

The eXtended Formal Ontology (XFO) Model Layer goes beyond BFO2 to incorporate states, state changes, schemas/frames, causation, and procedures. Moreover, as a rich semantic approach it is distinct from the weak semantics of simple linked data [8] that does not readily support pre-defined structured models or processes.

## 2 XFO DEVELOPMENT ENVIRONMENT

We developed a programming environment to implement XFO as a Model Layer over BFO2 and extensions to it. Here, we report the development of an XFO platform with Python, a widely-used language with support for several different programming paradigms. [1] Notably, for our purposes, Python

---

[1] [3] briefly explored using the Slate language [14] for modeling ontologies. Slate is regarded as implementing a relatively pure object-oriented model but is no longer supported.

supports prototypes via class copy, dynamic inheritance, code exec, and threads.

The XFO platform implemented an "Is_A" hierarchy of Entities, starting with "B" Entities as defined by BFO2. Applications included Universal ("U") Entities that are descended from B Entities or other U Entities. Particular ("P") Entities were mapped from U Entities with the "Instance_Of" relationship. Because every P Entity is descended from a corresponding U Entity, there was considerable redundancy between the two layers.

Pairs of defined U-Entities inherit the relationships of their B parents. For example, we say that Pottery Participates_In Biscuit Firing much the way that we would say that an object-oriented pottery class is associated with the biscuit firing method. It would also be acceptable to say that Pottery Participates_In Firing where Biscuit Firing is a specific type of Firing. By comparison, it does not normally make sense to say that Pottery Participates_In other activities such as Driving. Because of such considerations, we implemented extensive validation and checking of entities and relationships.

The status of Entities associated by Relationships with a BFO2 Object can be designated as their State. In addition to built-in Relationships (e.g., Participates_In), BFO2 also allows for ad hoc Relationships via Relational_Qualities. We term the ensemble of an Independent Continuant with its associated Relationships a Thick Independent Continuant (TIC). This is analogous to a class in object-oriented programming.

In BFO2, Processes are a type of Occurrent. Examples of Processes include the water flowing in a river, a person's running, or a person's lifetime. While state changes may be derived from Processes, BFO2 does not include any state changes.

State changes are needed for modeling Entities at the P-layer (Section 3). To allow for state changes, we define Transitionals as a model-level construct in addition to Processes. Such Transitionals make or break relationships between Entities. For instance, the change of the color of an object such as a Traffic Light would be modeled through unlinking one Color Quality (a Dependent Continuant) and linking another Color Quality. We may also define the operation of Entities in general with modeling at the U-Layer.

In addition to simple Objects, BFO2 allows for Object_Aggregates. BFO2 Object_Aggregates are Material Entities composed of parts that are not spatially unified. An orchestra is an Object_Aggregate. An Object_Aggregate typically would have a complex internal structure of Roles and interactions among the components (e.g., players in an orchestra). A composite with those internal structures would need to be defined in the Model Layer (see Section 4).

It is also useful to define procedures, which are combinations of Transitionals such as Mechanisms and Workflows, in the Model Layer.[2] Mechanisms [13] are collections of transitions that result in specific state changes. They are particularly important as causal explanations. A Workflow involves an external factor (often a person) to effectuate the transitions. The representation of Mechanisms and Workflows often includes control flow such as loops and conditionals.

Both Mechanisms and Workflows could be included as Occurrents with complex interactions among their components. For modeling, such interactions should be specified in detail. The specifications should confirm that a Mechanism or Workflow is complete in describing an end-to-end activity, at least within the limitations of the available terms. Moreover, each stage of the procedures (essentially Transitionals) should have its own metadata. When some stages are not known, there could be placeholders for the unidentified activities.

We allow TICs to interact in a Microworld. This approach is particularly well suited for an object-oriented and systems science paradigm. P-Layer

---

[2] [11] makes similar points in discussing alternatives for describing biomedical processes.

instantiations of Microworlds will need to allow for interrupted and broken procedures.

XFO incorporates a wide variety of checks on Entities, such as ensuring that terms are uniquely defined and that they are descended from a B entity. As is typical of programming environments, there is a tradeoff between checking validity when initially specifying the semantic constraints and checking at run-time [2].

Figure 1 is a code fragment from the XFO platform showing validation of a Continuant Part_Of Relationship. Currently, the XFO platform has about 800 lines of code for the base and an additional 800 lines for applications. It covers the main features described above but some details are not yet fully implemented.

```
elif(rName=='Continuant_Part_Of'):
    print('Continuant-Part_Of')
    eFromBparent = self.ecls[entFromId].BparentName
    eToBparent = self.ecls[entToId].BparentName
    if(entToName!='B_Continuant'):
        eFromBparentName = self.ecls[entFromId].BparentName
        testParentName = eFromBparentName
        testParentId = self.ec[testParentName]
```

Figure 1: Code fragment showing validation of a Continuant Part_Of Relationship.

## 3 EXAMPLE APPLICATION: TRAFFIC LIGHTS

As an initial demonstration of XFO, we implemented a Microworld with two simulated, asynchronous, Traffic Lights. Each of the Lights had three Lamps (green, yellow, red). Each Lamp entity is linked or unlinked to its corresponding Color Quality in a sequence specified in the program.[3]

In our implementation each Lamp is either "on" with the assigned color, or is "off". It could be argued that a Lamp has no color when it is "off", certainly not one of the standard Traffic Light colors. As an expedient, we defined the color of the "off" Lamps as "dark".

The behavior of the Lamps was implemented with imperative programming. The sequencing of the Lamps is controlled by P-level code. For example, durations for the Lamp color states were implemented with the Python sleep command. The separate Lights acted as independent objects. Python threads allowed the two Lights to operate asynchronously.

The interaction of objects in the Traffic Light Microworld would be highlighted if additional agents (e.g., cars, pedestrians, a power source) were added. It would be possible to develop a generic U-level procedure for Traffic Light sequencing, and that could be instantiated with specific values (e.g., Lamp duration).

Figure 2 shows two visualization panels that displayed the states of the Lights in the application. The upper panel show a real-time display while the lower panel shows cumulated states across the session. These were implemented with the Python Zelle graphics package.

---

[3] There are other ways in which the individual Traffic Lights and Lamps could have been programmed. For instance, rather than having Lamps of different colors, the Light itself could have been assigned three colors.

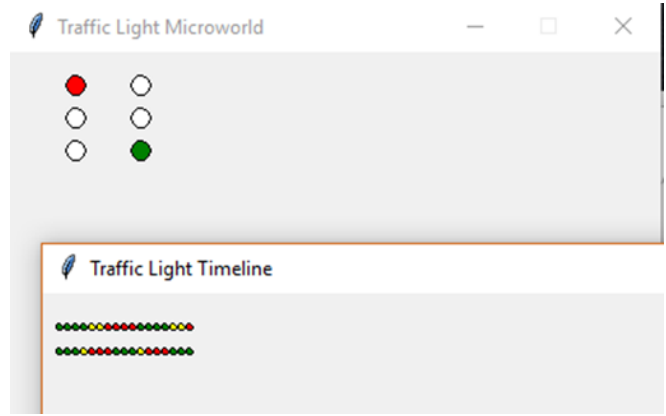

Figure 2: Two panels showing the state of the Traffic Light Microworld as it is being run. The upper panel shows the current state of the Lamps. The lower panel shows a temporal history of the Light changes on a timeline.

## 4 ADDITIONAL APPLICATIONS

We have begun to explore two more complex applications related to digital humanities, and addressed some issues in those applications. In the first of these applications, we modeled the activities of a School Superintendent from Nebraska from the first decade of the 20$^{th}$ Century [9]. Specifically, we modeled a news report that describes the Superintendent traveling to hire a new teacher to replace one who had resigned.

For the XFO program, we define the Norfolk School System as an Object_Aggregate. Superintendent and Teacher were Roles associated with the School System. Each of the Roles was described by an Employment frame (schema) which included slots for the Role, the Organization, the Person filling the Role, Compensation, Duration, Rights, Responsibilities, as well as associated actions (e.g., hire, fire, resign). This collection of linked Entities was implemented as a Frame and activated or deactivated as a unit (see [2]). We do not claim the Frame has semantic value beyond the sum of its parts.

One of the Responsibilities of the Superintendent was to replace teachers who resigned. There are several ways this model could be implemented. We used a condition-action pair that tested whether there was a teacher vacancy. If so, the Superintendent would select from among strategies for filling it. From the news report, we know that the Superintendent traveled by train find a replacement. Therefore, we modeled that action, and did not model alternatives or the decision processes in selecting among them.

In a second application, we developed an XFO program for celadon pottery production (see [5]). This program addressed two implementation issues. The first concerned the description of the raw clay that is used to create the pottery. Raw clay does not fall readily within the existing types of BFO2 Material Entities. Thus, we adopted the concept of Substance and classified raw clay as a Substance.

The second issue was the U-level pottery-making Workflow which we implement as a linked list of transitions. In addition to the metadata for the Workflow as a whole, each transition in the Workflow may have agent, duration, preconditions, and postconditions. Ultimately, the celadon-making Workflow could be embedded as part of a community Microworld which could include the kilns, workers, transportation of the pottery, etc.

## 5 OBJECT-ORIENTED PROPERTIES

The goal of ontology is to identify the basic Entities that describe the world. We propose extending BFO2 with XFO to develop models for how those Entities interact. As we have suggested, at least to a first approximation, XFO has many similarities to object-oriented analysis, design, and modeling. We believe that it is worthwhile to bring the two approaches closer together. Traditionally, object-oriented approaches are said to involve four factors: inheritance, abstraction, polymorphism, and encapsulation. It is clear that XFO supports inheritance and abstraction. Polymorphism asserts that there should be a single interface for interacting with Entities regardless of the datatypes. As we have observed (Section 2), an upper ontology defines a form of hierarchical semantic data typing. Encapsulation refers to an enclosed object containing data and functions that act on the enclosed data. While XFO does not currently implement "private" data, it would be relatively easy to add that feature.

The Bunge-Wand-Weber (BWW) ontology is sometimes considered as a more formal foundation for object-oriented modeling (e.g., [12]). We believe that BFO2/XFO provides a better approach.

# 6   DISCUSSION

XFO is a platform for developing a model layer, on top of BFO2. It includes features several such as Frames, Transitionals, and Workflows. The XFO platform should help us develop rich semantic descriptions that range from short abstracts to detailed representations. Text generation could be applied to the representations and incorporate discourse structures that go beyond rich semantics.

While we have focused on applications (e.g., descriptions of science and history) which people are generally willing to curate, ultimately, BFO2/XFO could provide a general framework for natural language processing. For instance, they could be an interlingua for translation. In addition, the structured representations of the Microworlds could be used for training artificial language users with machine learning [1, 11].

The U and P Entities used in these current applications were ad hoc, but eventually they should be drawn from standard model-level knowledgebases. Such model-level knowledgebases could be developed by incorporating sources such as BFO2 Foundries and faceted classification systems such as the Art and Architecture Thesaurus (AAT) or Bliss Classification (BC2).

# 7   REFERENCES

1   Allen, R.B. (1992) Connection Language Users, In N. Sharkey (ed.) Connectionist Natural Language Processing, 163-195.
2   Allen, R.B. (2014) Frame-based Models of Communities and their History. Histoinformatics 2013, LNCS 8359, 110-119, doi: 10.1007%2F978-3-642-55285-4_9
3   Allen, R.B. (2015) Repositories with Direct Representation, Networked Knowledge Organization Systems, arXiv: 1512.09070
4   Allen, R.B. (2017) Rich Semantic Models and Knowledgebases for Highly-Structured Scientific Communication, arXiv:1708.08423
5   Allen, R.B., & Kim, Y. (2017) Semantic Modeling with Foundries, arXiv: 1801.00725
6   Allen, R.B., Yang, E., & Timakum, T. (2017) A Foundry of Human Activities and Infrastructures, 2017, arXiv: 1711.01927, also ICADL 2017, LNCS 10647, doi: 10.1007/978-3-319-70232-2_5
7   Arp, R., Smith, B., & Spear, A.D. (2015) Building Ontologies with Basic Formal Ontology, MIT Press, Cambridge MA
8   Baker, T., & Sutton, S. (2015) Linked Data and the Charm of Weak Semantics: Introduction: The Strengths of Weak Semantics, ASIST Bulletin, 41(4), 10-12
9   Chu, Y.M., & Allen. R.B. (2016) Formal Representation of Socio-Legal Roles and Functions for the Description of History, TPDL, 2016, 379-385, doi: 10.1007/978-3-319-43997-6_30
10  Davidson, D. (2008) Time in Anatomy, In A. Burger, D. Davidson, and R Baldock (eds), Anatomy Ontology for Bioinformatics: Principles and Practice, Springer, 213-247
11  Hu, R., Andreas, J., Rohrbach, M., Darrell,T., & Saenko, K. (2017) Learning to Reason: End-to-End Module Networks for Visual Question Answering, arXiv: 1704.05526
12  Kiwelekar, A.W., & Joshi, R.K. (2007) An Object-Oriented Metamodel for Bunge-Wand-Weber Ontology, Workshop on Semantic Web for Collaborative Knowledge Acquisition, arXiv: 1004.3640

**13** Machamer, P., Darden, L., & Carver, C. (2000) Thinking about Mechanisms, Philosophy of Science, 67(1), 1-25.
**14** Salzman, L., & Aldrich, J. (2005) Prototypes with Multiple Dispatch: An Expressive and Dynamic Object, ECOOP, 312-336, doi:10.1007/11531142_14